\documentclass[journal,twoside,web]{ieeecolor}
\pdfoutput=1

\usepackage{tmi}
\usepackage{cite}
\usepackage[pdftex]{graphicx}
\graphicspath{{./figs/}}
\DeclareGraphicsExtensions{.pdf,.png,.jpg,.eps}
\usepackage{amsmath,amssymb,amsfonts}
\usepackage{mathrsfs}
\usepackage{pifont}
\usepackage{algorithmic}
\usepackage{array}
\usepackage{multirow}
\usepackage{textcomp}
\usepackage{url}
\usepackage{diagbox}
\usepackage{threeparttable}
\usepackage{makecell}
\usepackage{pgfplots}
\usepackage{comment}
\usepackage{booktabs}
\usepackage{stfloats}
\pgfplotsset{compat=1.18}

\newcommand{\cmark}{\ding{51}}
\newcommand{\xmark}{\ding{55}}

\def\BibTeX{{\rm B\kern-.05em{\sc i\kern-.025em b}\kern-.08em
    T\kern-.1667em\lower.7ex\hbox{E}\kern-.125emX}}

\begin{document}
\title{Cross-domain Denoising for Low-dose Multi-frame Spiral Computed Tomography}
\author{Yucheng Lu, Zhixin Xu, Moon Hyung Choi, Jimin Kim, and Seung-Won Jung, \IEEEmembership{Senior Member, IEEE}
\thanks{This work was supported by the Korea Medical Device Development Fund grant funded by the Korea government (the Ministry of Science and ICT, the Ministry of Trade, Industry, and Energy, the Ministry of Health \& Welfare, the Ministry of Food and Drug Safety) (Project Number: 1711139124, KMDF$\_$PR$\_$20200901$\_$0096). \emph{(Corresponding author: Seung-Won Jung.)}}
\thanks{Y. Lu is with the Education and Research Center for Socialware IT, Korea University, Seoul, Korea; and the Department of Datalogi, IT University of Copenhagen, Copenhagen, Denmark (e-mail: yucheng.l@outlook.com).}
\thanks{Z. Xu and S.-W. Jung is with the Department
of Electrical Engineering, Korea University, Seoul,
Korea (e-mail: hp19961109@korea.ac.kr; swjung83@korea.ac.kr).}
\thanks{M. H. Choi and J. Kim are with the Department of Radiology, Eunpyeong St. Mary's Hospital, College of Medicine, The Catholic University of Korea (e-mail: choimh1205@gmail.com; keithiix3926@gmail.com).}
}
\maketitle

\begin{abstract}
Computed tomography (CT) has been used worldwide as a non-invasive test to assist in diagnosis. However, the ionizing nature of X-ray exposure raises concerns about potential health risks such as cancer. The desire for lower radiation doses has driven researchers to improve reconstruction quality. Although previous studies on low-dose computed tomography (LDCT) denoising have demonstrated the effectiveness of learning-based methods, most were developed on the simulated data. However, the real-world scenario differs significantly from the simulation domain, especially when using the multi-slice spiral scanner geometry. This paper proposes a two-stage method for the commercially available multi-slice spiral CT scanners that better exploits the complete reconstruction pipeline for LDCT denoising across different domains. Our approach makes good use of the high redundancy of multi-slice projections and the volumetric reconstructions while leveraging the over-smoothing problem in conventional cascaded frameworks caused by aggressive denoising. The dedicated design also provides a more explicit interpretation of the data flow. Extensive experiments on various datasets showed that the proposed method could remove up to 70\% of noise without compromised spatial resolution, and subjective evaluations by two experienced radiologists further supported its superior performance against state-of-the-art methods in clinical practice. Code is available at \url{https://github.com/YCL92/TMD-LDCT}.
\end{abstract}


\section{Introduction}
\label{sec:introduction}
\IEEEPARstart{C}{omputed} tomography (CT) is one of the most popular tools used in clinical examinations nowadays due to its non-invasive and volumetric data acquisition advantages. Unlike conventional X-ray studies that project all volumetric information onto a single planar image, CT enables us to restore piles of axial structures through reverse reconstruction, providing rich space of data representation that helps access finer patterns for further evaluation and diagnosis.

Despite its great convenience and performance, there have been significant concerns about the potential health hazard to patients. Cell and organ damage may occur due to excessive exposure if proper measures against ionizing radiation are not considered. Even though the dosage of CT is relatively low, exposure over a protracted time can still increase the risk of developing cancer \cite{ortiz2021magnetic}. Therefore, minimizing radiation exposure has been carried out with a sense of urgency regarding science and public opinion~\cite{brenner2007computed}.

Since it is currently impractical to exclude CT from general health examinations, engineers have been working actively on reducing the radiation exposed to the subjects through various techniques, such as enlarging the source field of view, increasing the sensor resolution both horizontally and vertically, improving the detector sensitivity, and increasing the table speed for faster scanning. These efforts have led to considerable performance improvement over the past decades.

In addition to hardware-based solutions, researchers have also paid attention to dosage reduction through software-assisted technology. Two representative methods are sparse-view CT, which uses a reduced number of projections per gantry rotation, and low-dose CT (LDCT), which uses a reduced intensity of the X-ray source. Whereas the former tries to compensate for the artifacts introduced by the missing views, the latter gives rise to an image denoising problem, receiving more attention as similar tasks in other areas (e.g., low-light image enhancement) have been extensively studied \cite{buades2005non, dabov2007image, lu2022lowlight}.

With the success of convolutional neural networks (CNNs) in low-level computer vision tasks such as denoising \cite{abdelhamed2020ntire}, deblurring \cite{nah2021ntire}, and super-resolution \cite{cai2019ntire}, they have been rapidly adopted to medical imaging applications. Early studies have shown the promising potential of CNNs on LDCT denoising compared to conventional handcrafted regularizers \cite{jin2017deep, kang2017deep, chen2017low, wu2017iterative}. However, the protocol of LDCT denoising significantly differs from that of conventional image denoising, requiring further optimization of the obtained data representation (i.e., projections) before the final reconstruction. Unfortunately, many existing works directly borrow ideas from conventional image denoising and apply them to the reconstructed CT slices as a post-processing stage, which can be sub-optimal for LDCT denoising. Although there are a few works in the literature that handle projection data \cite{wurfl2016deep, yin2019domain, li2019learning, he2020radon, zhang2021clear}, they either operate on simulated 1D parallel-beam projection data which is not aligned with the modern CT scanner geometry, or try to incorporate the entire reconstruction pipeline into a single black-box model. All the above limitations hinder these works from being more optimal and transparent.

To address these problems, we propose a two-stage denoising framework dedicated to multi-slice spiral CT scanners in this paper. Specifically, in the first stage, a projection domain denoising network takes as input the successive projection slices and estimates sequential noise components, which are then rebined and used by the image restoration network in the second stage for further refinement. This two-stage design considers the domain-specific characteristics while avoiding information degradation in common cascaded structures, yielding objectively and subjectively more satisfactory image quality. In summary, the main contributions of this paper are as follows:
\begin{itemize}
    \item We propose a two-stage framework for LDCT denoising. The proposed method works across both the projection and image domains. It is specifically optimized for CT scanners with multi-slice helical geometry.
    
    \item We model each stage's physical properties of noise and artifacts based on the data acquisition process in the reconstruction pipeline. This design improves the denoising performance and gives end-users richer interpretation ability and transparency.
    
    \item We demonstrate through experiments on patient data that our method significantly outperforms existing works both quantitatively and qualitatively. An extensive analysis of phantom scans further supports that the proposed method has achieved state-of-the-art performance.
\end{itemize}

The remaining sections of this paper are organized as follows: Section II reviews some existing works related to our topic and briefly discusses their limitations, Section III presents the proposed method in full detail, Section IV provides the experiment setup and compares our results with those of several representative works, and finally, Section V concludes the paper.

\section{Related Work}
\label{sec:relatedworks}
Since the data acquisition and image restoration protocol of CT differs from that of conventional digital cameras, the design of an LDCT denoiser can be highly flexible depending on the appearance of the data to be processed. Compared to early works based on handcrafted image priors, such as total variation \cite{zhang2016statistical}, non-local means \cite{chen2009bayesian}, dictionary learning \cite{xu2012low}, block matching and 3D filtering \cite{feruglio2010block}, \emph{etc.}, some pioneering works \cite{jin2017deep, kang2017deep, chen2017low, wu2017iterative} have shown the superior potential of CNNs in LDCT denoising. Thus, we mainly focus on CNN-based methods and classify the most recent research into three categories: post-reconstruction image denoising, model-based iterative image reconstruction and denoising, and cross-domain joint optimization.

\subsection{Post-reconstruction Image Denoising} \label{sec:prev1}
Post-reconstruction image denoising aims at removing noise directly from the reconstructed CT images, which can be formulated as:

\begin{equation}
    I_{pr} = \mathcal{G}_{I} \left ( I_{n} \right ),
    \label{eq1}
\end{equation}

\noindent where $I_{n}$ and $I_{pr}$ are the low-dose noisy input and noise-suppressed output predicted by the denoiser $\mathcal{G}_{I}$, respectively. A significant merit of this approach is that it works in the 2D image domain as a post-processing step, so there is no need to alter the reconstruction pipeline in clinical practice. A considerable number of works fall within this category: Fan \emph{et al.} \cite{fan2019quadratic} replaced the conventional 2D convolution \cite{chen2017low} with the quadratic representation. Zavala \emph{et al.} \cite{zavala2022noise} revisited the perfect reconstruction conditions of the encoder-decoder-structured CNNs with soft-shrinkage and proposed a learnable shrinkage layer to handle decomposed wavelet frames, which was later extended using the over-complete Haar wavelet transform \cite{zavala2021image}. Liang \emph{et al.} \cite{liang2020edcnn} employed a perceptual loss from a pre-trained VGGNet to their densely connected denoiser. Matsuura \emph{et al.} \cite{matsuura2020feature} extracted context-aware features from images reconstructed with various parameter presets as data augmentation to enhance the quality of the denoised result. Tao \emph{et al.} \cite{tao2019vvbp} observed the specific patterns lying in the stacked view-by-view back-projection tensors and developed a tensor singular value decomposition-based algorithm for LDCT, which was extended to the field of deep learning-based LDCT denoising \cite{tao2021learning}. Xu \emph{et al.} \cite{xu2021low} applied dynamic filters predicted from a CNN to the extracted image features such that the non-uniformly distributed noise can be separated. Kim \emph{et al.} \cite{kim2021weakly} proposed a progressive denoising method to remove noise in an iterative manner while injecting synthetic noise into the projections on the fly. Zhang \emph{et al.} \cite{zhang2021transct} designed a framework consisting of two parallel networks to handle the low-frequency and high-frequency components, where the popular and powerful Transformer architecture \cite{vaswani2017attention} was adopted. Similarly, Wang \emph{et al.} \cite{wang2021ted} introduced reshaping, dilated unfolding, and cyclic shifting operations between successive Transformer blocks to share information across different patches, reaching better performance.

Unfortunately, clean observations can never be reached due to the statistical uncertainty of CT. Hence, the paired dataset used for training actually consists of routine-dose images, which serve as approximations of their noise-free counterparts, and low-dose images, which can be simulated via projection-domain noise injection or image-domain superimposition \cite{huber2022clinical}. As the noise statistics in CT images still follow some common properties such as zero-mean and zero discrepancies, a few researchers adopted the idea of Noise2Noise \cite{lehtinen2018noise2noise} and Noise2Void \cite{krull2019noise2void} in training their models without clean images: Zhang \emph{et al.} \cite{zhang2021noise2context} proposed using adjacent slices to approximate self-similarity, whereas Niu \emph{et al.} \cite{niu2022noise} further extended it to handle uncorrelated noise and structural artifacts with a broader range of patch-searching volumes. Alternatively, adversarial learning is also an option for unsupervised learning, which is usually achieved by generative adversarial networks (GANs) \cite{yi2019generative}: Shan \emph{et al.} \cite{shan20183} employed 3D convolution in the design of the encoder-decoder network and supervised the transfer learning from a pre-trained 2D variant using the Wasserstein distance. Yang \emph{et al.} \cite{yang2018low} combined the supervised and unsupervised learning with a hybrid loss term consisting of an adversarial loss and a perceptual loss to allow denoising while maintaining structural details, which was further improved by Li \emph{et al.} \cite{li2020sacnn} with the help of the self-attention module as well as the self-supervised perceptual loss. Ghahermani \emph{et al.} \cite{ghahremani2022adversarial} introduced an adversarial distortion learning method that considers the element-wise discrimination loss, reconstruction loss, pyramidal texture loss, and histogram loss in the supervision. Zhang \emph{et al.} \cite{zhang2021artifact} employed an artifact and noise attention network and used an edge feature extraction path to compensate for the over-smoothed details. Gu \emph{et al.} \cite{gu2021adain} adopted the cycle consistency and proposed a CycleGAN~\cite{zhu2017unpaired}-based model with adaptive instance normalization layers, achieving improved performance over the conventional CycleGAN at the cost of about only half of the parameters. Similarly, Lee \emph{et al.} \cite{lee2021iscl} introduced a pseudo network along with the CycleGAN framework and added a bypass consistency to prevent the generator from learning to embed blind information of noise into the output.

\subsection{Iterative Image Reconstruction and Denoising} \label{sec:prev2}
Although the post-reconstruction CT image denoising is simple and fast, an obvious limitation is that the high-pass filter (e.g., ramp filter) in the back-projection operation inevitably amplifies the noise component and introduces signal-dependent artifacts, which makes denoising more challenging and thus deteriorates the performance. To cope with this problem, model-based image reconstruction (MBIR) comes to the rescue, given by:

\begin{equation}
    I_{pr} = \mathop{\arg\min}_{I} \left\| \psi I - P_{n} \right\|_{2}^{2} + \lambda \phi \left ( I \right ),
    \label{eq2}
\end{equation}

\noindent where $\psi$ is the forward-projection operation that maps the reconstructed image $I$ back to the projection domain, $P_{n}$ represents the corresponding low-dose noisy observation, $\phi$ is a regularization function, and $\lambda$ is a balancing parameter.

The solution of \eqref{eq2} is typically obtained in an iterative manner that updates the reconstructed image by comparing the forward-projection result with the measurement under some constraints to stabilize the optimization. Many methods have been presented to embed pre-trained CNN denoisers as a part of the update protocol and achieved better performance against post-reconstruction denoising: Gupta \emph{et al.}, \cite{gupta2018cnn} utilized a CNN to project the objective function onto the data manifold and proposed a relaxed version of the projected gradient descent method that guarantees the convergence of the optimization. Kang \emph{et al.} \cite{kang2018deep} reviewed the denoising task under the low-rank Hankel structured matrix constraint and presented a wavelet residual network that learns to impose the low-rankness. Chen \emph{et al.} \cite{chen2018learn} proposed to replace the generalized regularization term referred to as the fields of experts \cite{roth2005fields} with a three-layer CNN, in which the trainable parameters are independent at each iteration. Similar work was presented by Aggarwal \emph{et al.} \cite{aggarwal2018modl} with their proposed conjugate gradient optimization-based data consistency layer, enabling the training of the unrolled model to be performed in an end-to-end manner with minimal memory cost. Inspired by \cite{chen2018learn}, Xia \emph{et al.} \cite{xia2021magic} further employed a learned graph convolutional network as an additional constraint to enhance non-local topological features in the low-dimensional patch manifold. He \emph{et al.} \cite{he2018optimizing} reformulated the problem as a dual-domain optimization task and modified the iteration of the alternating direction method of multipliers (ADMM) by using CNNs to represent the gradient, resulting in a parameterized plug-and-play ADMM optimization scheme. Chun \emph{et al.} \cite{chun2019bcd} introduced BCD-Net \cite{chun2018deep} to LDCT reconstruction and applied the accelerated proximal gradient method as a fast numerical solver. Ye \emph{et al.} \cite{ye2021unified} took both the supervised regularization and the unsupervised regularization into account and proposed an optimization scheme to alternatively update the reconstruction result under specific constraints, where the experiments on several publicly available MBIR-based methods showed improved performance against their vanilla counterparts.

\subsection{Cross-domain Joint Optimization} \label{sec:prev3}
MBIR methods generally yield higher reconstruction quality. However, they significantly increase the reconstruction time and typically occupy more computational resources. To achieve fast inference speed while reserving access to raw projection data, a promising solution is to apply denoising across different data domains that cover the complete reconstruction pipeline, as follows:

\begin{equation}
    I_{pr} = \mathcal{G}_{I} \left ( \varphi \left ( \mathcal{G}_{P} \left ( P_{n} \right ) \right ) \right ),
    \label{eq3}
\end{equation}

\noindent where $\mathcal{G}_{P}$ and $\mathcal{G}_{I}$ are denoisers in the projection domain and image domain, respectively. $\varphi$ is a projection-to-image operation that can be performed by conventional algorithms such as filtered back-projection (FBP) or learned models.

Several dedicated works fall into this category: Würfl \emph{et al.} \cite{wurfl2016deep} utilized a multi-layer perceptron to model the behavior of filtering, back-projection, and non-negative constraint for sinogram-to-image reconstruction. Li \emph{et al.} \cite{li2019learning} designed a model named iCT-Net consisting of a novel back-projection layer for both LDCT denoising and sparse-view reconstruction. He \emph{et al.} \cite{he2020radon} presented a deep learning-based Radon inversion framework, where the filtered sinogram is resampled by a sinusoidal back-projection layer, followed by a typical CNN. Zhang \emph{et al.} \cite{zang2021intratomo} adopted the Fourier feature representation \cite{tancik2020fourier} in their proposed sinogram prediction module and designed an iterative optimization scheme via forward and backward projection. Another work \cite{zhang2021clear} combined two 3D residual U-Nets (ResUNets) for the projection domain and image domain denoising and trained them using cross-domain supervision and adversarial learning, demonstrating state-of-the-art performance.

The methods above have contributed to LDCT denoising to some extent; however, there is room for improvement due to their limitations. For image domain denoising, lacking direct access to the projection data increases the difficulty of distinguishing between subtle structures and signal-dependent artifacts. For iterative image reconstruction and denoising, the simulated scanner, i.e., the planar scanning with parallel trajectory, is not aligned with multi-slice CT scanners rotating in helical mode. Applying to fit the actual geometry will likely introduce extra complexity to the optimization or even affect model convergence. As to cross-domain methods, not only a single image reconstruction for modern scanners typically requires thousands of projections to complete, which makes the training of $\mathcal{G}_{P}$ and $\mathcal{G}_{I}$ heavily unbalanced, but also some 3D operations in the reconstruction, such as the Feldkamp-like weighted FBP \cite{stierstorfer2004weighted}, are difficult to be replaced by learned models. Consequently, attempts at end-to-end optimization become even more challenging, whereas these cascaded models can be easily affected by the over-smoothing problem due to the aggressive denoising of separately trained sub-networks.

Unlike existing works, our proposed framework takes the complete characteristics of the CT image reconstruction pipeline into account and performs joint projection-domain denoising and image-domain refinement while avoiding the complexity and difficulty of end-to-end fine-tuning without compromised performance observed in other cascaded designs. To the best of our knowledge, there are only very few works closely related to ours in the literature: Yin \emph{et al.} \cite{yin2019domain} proposed to use two 3D sub-networks for sinogram and image denoising, respectively, where each sub-network was trained separately to take as input volumetric frames and estimate noise using 3D convolutions. However, this cascaded design still struggles to recover from aggressive denoising. Also, the effect of rebinning has not been considered. As a result, performance improvement is limited. In comparison, our method decomposes the reconstruction pipeline into several learning-based optimization problems according to the characteristics of data representation, which yields higher transparency to clients. This two-stage design not only improves the overall performance by a significant margin but also strengthens the system's robustness subject to different gantry geometries.

\section{Proposed Method}
\label{sec:proposedmethod}
In this section, we first discuss the intuition behind the design to offer the readers a brief overview and then provide more details of the proposed framework.

\begin{figure*}[htb]
    \centering
    \includegraphics[scale=1]{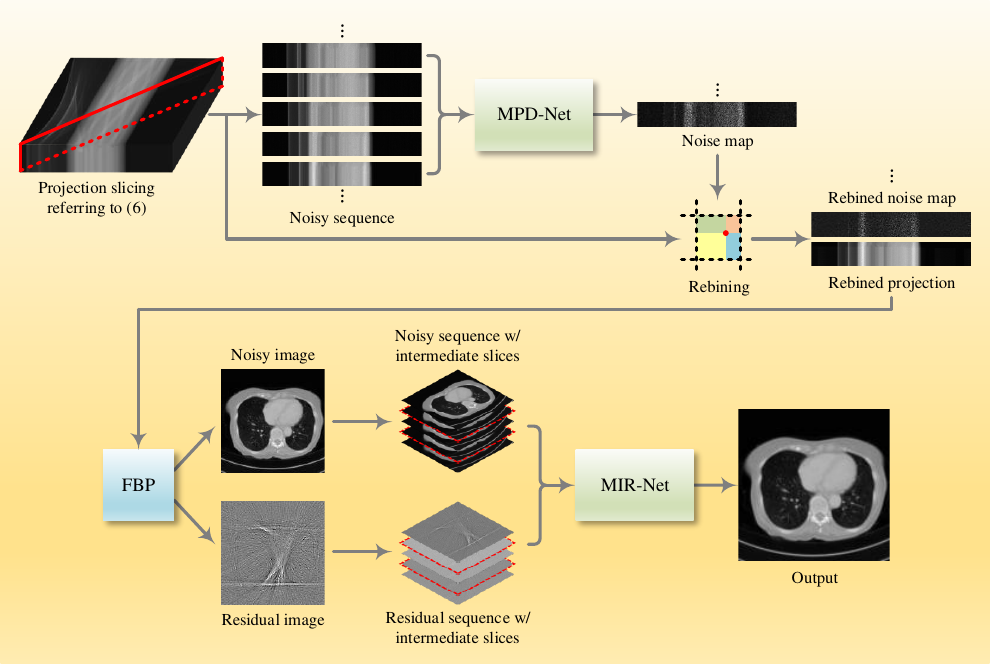}
    \caption{Overview of the proposed multi-stage hierarchical framework. The curly brackets indicate the concatenation operation. Note that only a single stream is shown in the projection-domain denoising, and all the noise components are amplified for better visibility.}
    \label{fig-overview}
\end{figure*}

\subsection{Reconstruction Revisit}\label{sec:revisit}
Pixels in a CT projection are obtained through the line integrals along the attenuating path. Without considering other effects (e.g., beam hardening), the ideal clean measurement $p_{c}$ is given as:

\begin{equation}
    p_{c} = -\ln \left ( \frac{N}{N_{0}} \right ),
    \label{eq4}
\end{equation}
where $N_{0}$ and $N$ are the incident and received intensity, respectively. However, noise is inevitable due to the quantum effects of photons. According to \cite{whiting2006properties}, a measurement under clinical environment, denoted as $p_{n}$, can be approximated by:

\begin{equation}
    p_{n} = p_{c} + \frac{x}{\sqrt{N_{0} \exp \left ( -p_{c} \right )}},
    \label{eq5}
\end{equation}
where $x \sim \mathcal{N} \left( 0,1 \right)$ is a unit Gaussian random variable. Similar to the light field camera \cite{ng2005light}, two extra variables further parameterize each measurement in the detector array to represent the spatial information of the ray, namely the ray distance $d$ to the isocenter and the ray angle $\alpha$ to the table. Hence, a complete ray representation is given as $p_{n} = P_{n}\left ( d,\alpha \right )$. Note that this process is general enough regardless of the source geometry (e.g., fan-beam or parallel-beam).

Modern CT scanners usually operate in helical trajectories, which requires an additional step called rebinning to convert raw projections to pseudo-parallel geometry via the following slicing operation:

\begin{equation}
    \hat{P}_{n} \left ( i \right ) = P_{n} \left ( d_{i},\alpha_{i} \right ),
    \label{eq6}
\end{equation}
where $i$ is an element index in the rebined projection data. As the slicing indices are usually fractions, this operation is basically a 2D interpolation.

After that, all the resulting projections are transformed into image representation through CT reconstruction methods, where FBP is a popular choice, simplified as:

\begin{equation}
    I_{n} \left ( u,v \right ) = \frac{\pi }{M}\sum_{m=1}^{M} \mathcal{\hat{P}}_{n} \left ( u \cos \theta_{m} + v \sin \theta_{m}\right ),
    \label{eq7}
\end{equation}
where $(u,v)$ denotes the image pixel location, $\mathcal{\hat{P}}$ represents the filtered result of $\hat{P}$, $\theta_{m}$ is the $m$-th projection angle, and $M$ is the number of rebined projections. Readers are referred to \cite{hsieh2003computed} for a more comprehensive understanding of image reconstruction on multi-slice spiral CT.

\subsection{Framework Overview}
An overview of the proposed framework is presented in Fig. \ref{fig-overview}. It mainly performs the projection-domain denoising and the image-domain refinement, embodied by two multi-frame-based neural networks, named MPD-Net and MIR-Net, respectively.

Let $S_{n*}^{P} = \left [ P_{n*}^{1}, P_{n*}^{2}, \cdots, P_{n*}^{K} \right ]$ denote a sequence of $K$ consecutive noisy projections, where $* \in \left \{ l,r \right \}$ denotes the left or the right candidate to be sampled in \eqref{eq6}, and the upper-script represents the time step, which can be omitted when unnecessary. Given $S_{n*}^{P}$ as the input, MPD-Net performs multi-frame noise estimation using a sliding window of size $2F+1$ for every projection, resulting in a denoised sequence $S_{d*}^{P} = \left [ P_{d*}^{F+1}, P_{d*}^{F+2}, \cdots, P_{d*}^{K-F} \right ]$ with a noise level similar to routine dose (i.e., full dose), as follows:

\begin{equation}
\begin{aligned}
    P_{d*}^{t} &= P_{n*}^{t} + R_{*}^{t}\\
    &= P_{n*}^{t} + \mathcal{G}_{MPD} \left ( P_{n*}^{t-F}, \cdots, P_{n*}^{t}, \cdots, P_{n*}^{t+F} \right ),
    \label{eq8}
\end{aligned}
\end{equation}
where $\mathcal{G}_{MPD}$ represents MPD-Net, $R_{*}^{t}$ represents its output, $t \in \left \{ F+1, F+2, \cdots, K-F \right \}$, and $K \gg F$.

The rebined projection sequence, denoted as $\hat{S}_{d}^{P} = \left [ \hat{P}_{d}^{F+1}, \hat{P}_{d}^{F+2}, \cdots, \hat{P}_{d}^{K-F} \right ]$, is then obtained by the weighted summation over adjacent projections as follows:

\begin{equation}
    \hat{P}_{d}^{t} = \omega_{00} P_{dl}^{t} + \omega_{01} P_{dr}^{t} + \omega_{10} P_{dl}^{t+1} + \omega_{11} P_{dr}^{t+1},
    \label{eq9}
\end{equation}

\noindent where $\omega_{00}$, $\omega_{01}$, $\omega_{10}$, and $\omega_{11}$ are interpolation weights that add up to one.

After that, the conjugate projections with angles $\tilde{\theta}_{j} = \theta_{j} + k\pi$ are filtered and back-projected onto the orthogonal 2D plane, forming a back-projection view $V_{\theta_{j}}$ as follows:

\begin{equation}
    V_{\theta_{j}}^{z}\left ( u, v \right ) = \frac{1}{H_{\theta_{j}}} \sum_{k}h\left ( z_{j,k} - z \right ) \mathcal{\hat{P}}_{d} \left ( u\cos\theta_{j} + v\sin\theta_{j} \right ),
    \label{eq10}
\end{equation}
where $z_{j,k} - z$ is the axial offset of the ray to the reconstruction center $z$, $h$ is a non-linear weighting function related to the multi-slice spiral geometry, and $H_{\theta_{j}}$ is the sum of $h$ over $k$.

A complete reconstruction can then be derived once a half-turn $L$ is reached:

\begin{equation}
    I_{d}^{z}\left ( u,v \right ) = \frac{\pi}{L}\sum_{j=1}^{L}V_{\theta_{j}}^{z}\left ( u, v \right ).
    \label{eq11}
\end{equation}

Similar to the projection domain denoising, when a sequence of $Q$ reconstructed images is collected, MIR-Net takes the sequence $S_{d}^{I} = \left [ I_{d}^{z_{1}}, I_{d}^{z_{2}}, \cdots, I_{d}^{z_{Q}} \right ]$ as input and generates the refined image as the final result using a sliding window of size $2F + 1$, given as:

\begin{equation}
    \begin{aligned}
        I_{r}^{z
    _{q}} &= I_{d}^{z_{q}} + R_{r}^{z_{q}}\\
        &= I_{d}^{z_{q}} +\mathcal{G}_{MIR}\left ( I_{d}^{z_{q-F}},\cdots,I_{d}^{z_{q}},\cdots,I_{d}^{z_{q+F}} \right )
        \label{eq12}
    \end{aligned}
\end{equation}
where $\mathcal{G}_{MIR}$ represents MIR-Net, $R_{r}^{z_{q}}$ corresponds to its output, and $q\in \left\{ F+1,F+2,\cdots,Q-F\right\}$. The input sequence to MIR-Net has a stride of $F + 1$, i.e., $q\left ( \mod\left ( F+1 \right ) \right ) \equiv 1$, which will be explained in Section \ref{sec:mirnet}.

\begin{figure}[htb]
    \centering
    \includegraphics[scale=1]{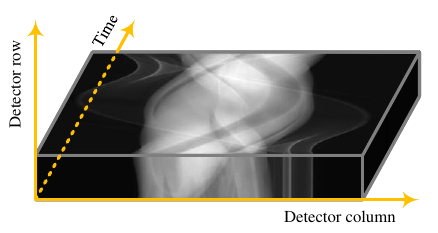}
    \caption{A sample clip from the raw projections obtained by a multi-slice spiral CT scanner.}
    \label{fig-redundancy}
\end{figure}

\subsection{Multi-frame-based Projection Denoising}
\label{sec:mpdnet}

We observe that modern scanners with multi-slice detectors provide intra-view (i.e., detector array) and inter-view (i.e., neighboring views) redundancy that are both beneficial for denoising. As can be seen from the example in Fig. \ref{fig-redundancy}, the intra-view similarity provides the projected structural details of the objects, and the inter-view similarity presents the relative motion between objects. As discussed in Section \ref{sec:revisit}, the dominant source of noise is the photon noise that follows a Poisson distribution, which can be alleviated by averaging over multiple independent measurements. We thus consider reformulating the task as a burst imaging problem and propose MPD-Net to capture both the intra-view and inter-view features. It features two significant merits: On the one hand, noise reduction based on the statistics of burst imaging can be realized via implicit alignment and fusion along views; on the other hand, it also considers intra-view similarities so that structures within a view can be well preserved.

\begin{figure}[htb]
    \centering
    \includegraphics[scale=1]{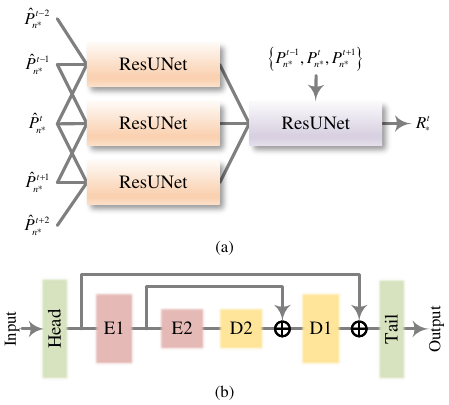}
    \caption{ The structure of MPD-Net: (a) MPD-Net with $F$=2, where ResUNets with different colors have individual parameter sets; (b) the structure of ResUNet. Details of E1-E2 and D1-D2 are given in Fig. \ref{fig-mirnet}.}
    \label{fig-mpdnet}
\end{figure}

The main structure of MPD-Net is depicted in Fig. \ref{fig-mpdnet}. Inspired by \cite{tassano2020fastdvdnet}, a multi-step denoising model consisting of two modified ResUNets is adopted. Each ResUNet takes $2F+1$ frames as input and predicts a denoised version of the middle frame via residual learning. Different from the original, where the same noise prior is used at each step \cite{tassano2020fastdvdnet}, we take the predicted residual from the first step along with the untouched frames as input of the second step to avoid potential accumulated artifacts. Furthermore, the adaptive mix-up from \cite{wu2021contrastive} is also employed empirically in boosting performance.

The above intuition based on multi-frame denoising could be sub-optimal since the rebinning operation defined in \eqref{eq6} generates pseudo-parallel projections obtained across time. In addition, some advanced scanners apply the flying focal spot technique to improve axial resolution, where the rebinning process comes with another step that interleaves rows from two focal spots. All these operations lead to inconsistent correlations along detector channels, as we will demonstrate in Section \ref{sec:ablation}, applying rebinning after MPD-Net results in unsmooth projections due to the absence of long-term consistency. However, if rebinning is performed ahead of MPD-Net, as it is essentially an interpolation operation, the element-wise noise independence will be violated, which degrades the denoising performance, as reported by \cite{brooks2019unprocessing, bao2020real}.

To guarantee element-wise noise independence while preserving rebinning consistency, we decompose the rebinning process into two steps, namely integer slicing and weighted summation, and bridge them with MPD-Net. The integer slicing extracts four untouched neighbors from 2D raw projections, whereas the weighted summation performs rebinning to the denoised results. This modification not only retains both the noise property and intra-view smoothness but also avoids complicated long-term memory mechanisms in the design.

\begin{figure}[htb]
    \centering
    \includegraphics[scale=1]{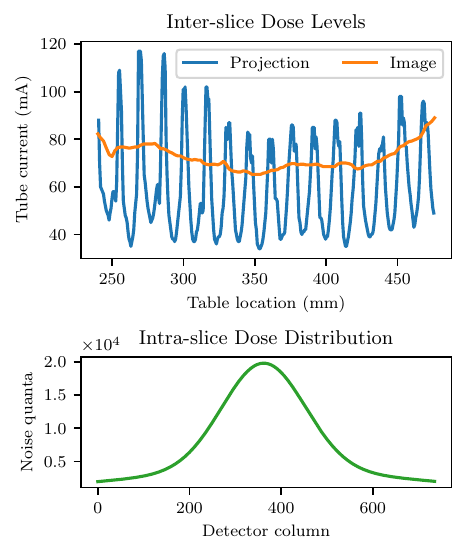}
    \caption{Two examples of inter-slice (study: Siemens-L291) and intra-slice (study: Siemens-L291, slice: 9,500) dose levels. It can be seen that the dose level not only varies dramatically within the scan but also has a non-uniform distribution across detector columns.}
    \label{fig-noise}
\end{figure}

Besides, as shown in Fig. \ref{fig-noise}, due to the wildly applied Automatic Exposure Control (AEC) \cite{hsieh2003computed}, not only the exposure levels of projections oscillate dramatically among table locations compared to that of images, but also the noise distributes non-uniformly within each projection. Although an existing work \cite{zeng2022performance} demonstrated a certain level of tolerance of CNNs against dose levels in the image domain, blind denoisers tend to learn a more aggressive strategy when the noise variance becomes large, resulting in more blurry predictions \cite{xu2018external}. Hence, we provide an external noise prior and train MPD-Net to recover the refined noise map. Given two sets of projections under low dose (dubbed $P_{l}$) and target full dose (dubbed $P_{f}$), let $N_{l_{0}}$ and $N_{f_{0}}$ be their corresponding source intensity. Considering a realization of $x$ in \eqref{eq5} be $X$, then $P_{l}$ can be approximated using $P_{f}$ by:

\begin{equation}
     P_{l}=P_{f} + \sqrt{\left ( \frac{1}{N_{l_{0}}} - \frac{1}{N_{f_{0}}} \right ) \exp\left ( P_{f} \right )}X.
    \label{eq13}
\end{equation}
Replacing $ P_{l}$ by $P_{f} + \Delta_{P}$, \eqref{eq13} can be rewritten as:

\begin{equation}
\begin{aligned}
    P_{f} &= P_{l} - \sqrt{\left ( \frac{1}{N_{l_{0}}} - \frac{1}{N_{f_{0}}} \right ) \exp\left ( P_{l} \right )} \cdot \sqrt{\exp\left ( -\Delta_{P} \right )}X\\
    &= P_{l} - \Phi \cdot \omega.
    \label{eq14}
\end{aligned}
\end{equation}
Interestingly, the second part in \eqref{eq14} can be viewed as the multiplication of a constant term $\Phi$, which we define as the \textit{noise prior}, and a weighting term $\omega$, which we define as the \textit{weight map}. Although noise-irrelevant, $\Phi$ keeps changing among views when AEC is enabled. As a result, noise estimation becomes more challenging due to the extra uncertainty from $\Phi$. Fortunately, the per-channel source intensity $N_{0}$ is required to compute the attenuation ratio in \eqref{eq4}. To alleviate the difficulty of noise estimation, we thus feed this noise prior to MPD-Net to help predict a more accurate noise map. The updated workflow defined in \eqref{eq8} is then given by:

\begin{equation}
\begin{aligned}
    P_{d*}^{t} &= P_{n*}^{t} + \mathcal{G}_{MPD} \left ( \hat{P}_{n*}^{t-F},\cdots,\hat{P}_{n*}^{t},\cdots,\hat{P}_{n*}^{t+F} \right ),
    \label{eq15}
\end{aligned}
\end{equation}
where $\hat{P}_{n*}^{t} = \left [ P_{n*}^{t}, \Phi_{*}^{t} \right ]$ is the concatenation of the noisy projection $P_{n*}^{t}$ and the corresponding noise prior $\Phi_{*}^{t}$.

\subsection{Multi-frame-based Image Refinement}
\label{sec:mirnet}
Although the proposed MPD-Net can significantly reduce the noise of multi-slice projections, it is still far from satisfactory for two reasons: First, MPD-Net does not capture the structural features of the final reconstructed image because the reconstruction plane is in parallel with the ray trajectories. Second, the remaining noise in the results will be amplified by the high-pass filter and lead to streak artifacts after reconstruction. To tackle these problems, we introduce a second network called MIR-Net to refine the LDCT image further.

\begin{figure}[htb]
    \centering
    \includegraphics[scale=1]{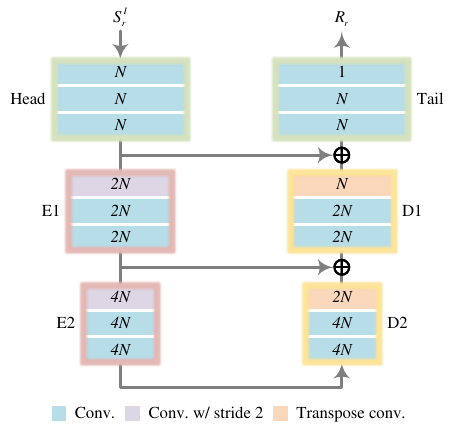}
    \caption{Structure of MIR-Net. The number in each layer represents the output channel size, and $\bigoplus$ represents the adaptive mix-up operation. E1-E2 and D1-D2 represent encoding and decoding blocks, respectively. All layers use ReLU as the activation function.}
    \label{fig-mirnet}
\end{figure}

Fig. \ref{fig-mirnet} presents the structure of MIR-Net, which consists of a single ResUNet. MIR-Net takes a reconstructed image sequence $S_{r}^{I}$ as input and produces the residual $R_r$ as the output. The hourglass design enables an expanding receptive field that better captures structural features without large kernels, where we observe more excellent performance against other straight (i.e., without down/up-sampling) networks. 

\begin{figure}[htb]
    \centering
    \includegraphics[scale=1]{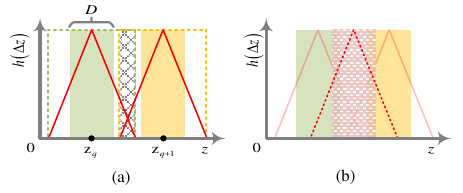}
    \caption{Illustration of slice relations: (a) a pair of non-overlapped slices when $\Delta z > D$; the dotted boxes are their projection data grabbing range, the red lines are corresponding weights that contribute to $h$ (w/o $w\left ( r \right )$), and the grid in between indicates shared projections; (b) the proposed multi-slice redundancy ($F$=1).}
    \label{fig-recon}
\end{figure}

Although MPD-Net benefits from multi-frame processing, it is challenging for MIR-Net to discover the full potential of multi-frame input. Let $I^{z_{q}}$ and $I^{z_{q+1}}$ be two consecutive CT images, and $D$ be the slice thickness. As can be seen in Fig. \ref{fig-recon}(a), when $\left| z_{q+1} - z_{q} \right| \geq D$, the (ideal) reconstructed CT images do not share objects between slices, meaning that the multi-frame input only features structural similarity instead of redundant observations. Besides, a 2D CT reconstruction represents a 3D volume in reality. Recalling \eqref{eq10} and \eqref{eq11}, artifacts from conjugate projections are combined in each view. In contrast, a stack of views over a half-turn is projected onto the 2D image plane, resulting in compressed artifact patterns that are more difficult to remove.

We propose a simple yet effective approach to alleviate these problems by introducing overlapped slices as intermediate representations. According to \eqref{eq10}, the reconstruction is obtained by averaging nearby back-projections through a weighting function $h$. We refer to the reconstruction method in \cite{stierstorfer2004weighted}, where $h$ is given as:

\begin{equation}
    \begin{aligned}
        h\left ( \Delta z \right ) = \max\left ( 0, 1 - \frac{\left| \Delta z \right|}{D} \right )w\left ( r \right ),
        \label{eq16}
    \end{aligned}
\end{equation}
where $r$ denotes the detector row index, and $\Delta z = \left| z_{q+1} - z_{q} \right|$ represents the distance between the projection and the reconstruction center along the table direction.

Without considering $w\left ( r \right )$, the amount of projection data required for a complete reconstruction, defined by $h\left ( \Delta z \right ) > 0$, simply lies in $\left ( z-D, z+D \right )$, indicating that the expansion of projections is wider than the slice. In other words, there may still be a small number of shared projections used in reconstructing both $I^{z}$ and $I^{z+1}$. These shared features provide redundant observations that are beneficial for multi-frame-based refinement. However, as Fig. \ref{fig-recon}(a) shows, the weights of shared projections become insignificant as $\Delta z$ increases; one could insert more slices in between to emphasize these weak signals better. To this end, we reconstruct $F$ slices as intermediate observations between each pair of adjacent slices and collect $2F + 1$ images as the multi-frame input. Fig. \ref{fig-recon}(b) presents the proposed solution when $F = 1$; by doing so, the emphasized projections offer a different realization of the compressed artifacts. It is worth mentioning that these intermediate representations cannot be obtained via interpolation from adjacent slices as the complete form of the weighting function is non-linear. Also, the shared features among adjacent slices are pixel-wise aligned; thus, a single ResUNet is efficient enough to handle multi-frame inputs.

Besides, we notice a certain degree of degradation in high-frequencies due to the aggressive denoising in individual sub-networks, which commonly occurs in cascaded designs. A straightforward solution is to fine-tune the entire chain in an end-to-end manner. However, as discussed in Section \ref{sec:relatedworks}, end-to-end training is expensive and impractical for multi-slice spiral CT. Alternatively, instead of obtaining the denoised results from MPD-Net, we use their concatenation form, i.e., $\left [ P_{n*}^{t}, R_{*}^{t} \right ]$. This not only compensates for the missing high-frequencies but also provides a decoupled reference of structural artifacts for further refinement. In short, the image domain refinement defined in \eqref{eq12} is rewritten as:

\begin{equation}
        I_{r}^{z_{q}} = I_{n}^{z_{q}} +\mathcal{G}_{MIR}\left ( \hat{I}_{d}^{_{q-F}},\cdots,\hat{I}_{d}^{z_{q}},\cdots,\hat{I}_{d}^{z_{q+F}} \right ),
        \label{eq17}
\end{equation}
where $\hat{I}^{z_{q}}_{d} = \left [ I^{z_{q}}_{n}, R^{z_{q}}_{n} \right ]$ is the concatenation of the low-dose image $I^{z_{q}}_{n}$ and the residual $R^{z_{q}}_{n}$ reconstructed using the raw projections and MPD-Net predictions, respectively.

\begin{table*}[b]
    \renewcommand{\arraystretch}{1.3}
    \caption{Dataset Partition Summary}
    \label{table-dataset}
    \centering
    \begin{tabular}{ccc}
        \hline
        Partition & Siemens Scanner Subset & GE Scanner Subset\\
        \hline
        \multirow{4}{*}{Training} & L004, L006, L019, L033, L057, L064, L071, L072, L081, L107 & L012, L024, L027, L030, L036, L044, L045, L048, L079, L082\\
        & L110, L114, L116, L125, L131, L134, L150, L160, L170, L175 & L094, L111, L113, L121, L127, L129, L133, L136, L138, L143\\
        & L178, L179, L193, L203, L210, L212, L220, L221, L232, L237 & L147, L154, L163, L166, L171, L172, L181, L183, L185, L188\\
        & L248, L273, L299 & L196, L216\\
        \hline
        Validation & L077, L148, L229 & L043, L213, L238\\
        \hline
        \multirow{2}{*}{Testing} & L014, L056, L058, L075, L123, L145, L186, L187, L209, L219 & L218, L228, L231, L234, L235, L244, L250, L251, L257, L260\\
        & L241, L266, L277 & L267, L269, L288\\
        \hline
    \end{tabular}
\end{table*}

\section{Experiments and Analysis}
\label{sec:experiments}
In this section, we present evaluation results and analysis of the proposed framework. We first provide implementation details and ablation studies to verify our design. We then evaluate its performance both quantitatively and qualitatively.

\subsection{Implementation Details}
The proposed framework was implemented in PyTorch and trained on an RTX 3090 GPU with i9-10980XE CPU and 128GB RAM. We chose Adam Optimizer to update the model parameters. The Low Dose CT Image and Projection Data V6 \cite{moen2021low}, obtained from two scanners (Siemens SOMATOM Definition Flash, dubbed Siemens dataset, and GE Discovery CT750i, dubbed GE dataset), was used for training, validation, and testing. Both Siemens and GE datasets contain paired full-dose (7.6-28.8 mGy for Siemens studies and 9.2-21.6 mGy for GE studies) and quarter-dose (1.9-7.2 mGy for Siemens studies and 2.3-5.4 mGy for GE studies) data. The detailed data partitions are given in Table \ref{table-dataset}. The test dataset of the 2016 Low-dose CT AAPM Grand Challenge \cite{ldct2016} was also employed in subjective evaluation since it contains scans from the control group, in which the dose level ranges from 0.8 mGy to 4.5 mGy. We followed the methods described in \cite{flohr2005image, stierstorfer2004weighted, hoffman2018free} for rebinning, filtering, back-projection, and weighted summation. The default Shepp–Logan filter was chosen as the reconstruction kernel, and the slice thickness from metadata was adopted.

During the network training, we used L1 loss to supervise both MPD-Net and MIR-Net. The initial learning rate was set to $1 \times 10^{-4}$, then reduced to $1 \times 10^{-5}$ if the model performance on the validation dataset had no further improvement after certain steps. The complete convergence of the two CNNs took about 40 and 150 epochs, respectively. For both networks, we set $F=2$ as the size of the sliding window.

\subsection{Ablation Studies}
\label{sec:ablation}
We conducted experiments to show how the image quality is progressively improved by each component in the proposed framework, namely MPD-Net, MIR-Net, multi-frame input, external noise prior, and decoupled input (i.e., separated $P_{n*}^{t}$ and $R_{*}^{t}$). Ablation studies were performed on the Siemens test dataset; we report the measured mean square error (MSE) and structural similarity (SSIM) \cite{wang2004image} in Table \ref{table-abla}.

\begin{table}[htb]
    \renewcommand{\arraystretch}{1.3}
    \caption{Results of Ablation Studies}
    \label{table-abla}
    \centering
    \begin{tabular}{ccccccc}
        \hline
        MPD & MIR &  Multi & Prior & Decoupled & MSE$\downarrow$ & SSIM$\uparrow$\\
        \hline
        \xmark & \xmark & \xmark & \xmark & \xmark & 774.46 & 0.9577\\
        \cmark & \xmark & \xmark & \xmark & \xmark & 499.93 & 0.9685\\
        \cmark & \cmark & \xmark & \xmark & \xmark & 418.54 & 0.9730\\
        \cmark & \cmark & \cmark & \xmark & \xmark & 309.79 & 0.9774\\
        \cmark & \cmark & \cmark & \cmark & \xmark & 236.60 & 0.9819\\
        \cmark & \cmark & \cmark & \cmark & \cmark & 184.98 & 0.9881\\
        \hline
    \end{tabular}
\end{table}

\begin{figure}[htb]
    \centering
    \includegraphics[scale=1]{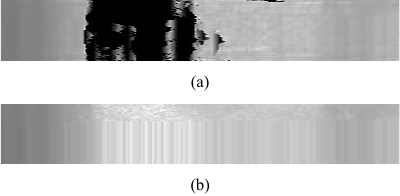}
    \caption{The interaction between the rebinning process and MPD-Net: (a) the result obtained from MPD-Net placed before the rebinning operation; (b) the result of the proposed method. Images are contrast-enhanced for better visibility.}
    \label{fig-rebin}
\end{figure}

It can be seen that both MPD-Net and MIR-Net play essential roles in enhancing the reconstruction quality of LDCT. For MPD-Net, introducing external noise prior leads to quality improvement, whereas placing the model before the rebinning operation results in sub-optimal performance, where we observe artifacts as shown in Fig. \ref{fig-rebin}. Similarly, it is noticed that the summation (i.e., $P_{n*}^{t} + R_{*}^{t}$) in \eqref{eq8} affects the reconstruction, mainly due to over-smoothing, whereas volumetric input brings extra improvement to the final image quality.

\begin{figure}[htb]
    \centering
    \includegraphics[scale=1]{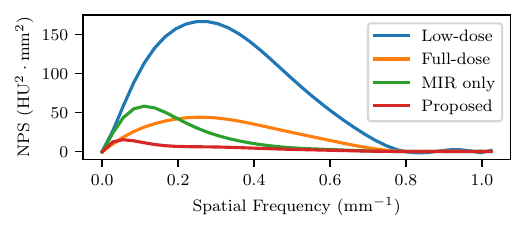}
    \caption{NPS results of ablation study on ACR CT phantom.}
    \label{fig-nps-abla}
\end{figure}

We also tested the performance of MIR-Net as a standalone image-domain denoiser. Although the results regarding MSE (215.62) and SSIM (0.9807) seem promising, further inspection of the noise power spectrum (NPS), shown in Fig. \ref{fig-nps-abla}, indicates an obvious over-smoothing problem evidenced by the shifted peak with a higher level of noise power in low frequencies. In short, all these results align well with our analysis above.

\begin{table*}[b]
    \renewcommand{\arraystretch}{1.3}
    \caption{Objective Evaluation Summary (Quarter-dose)}
    \label{table-objeval1}
    \centering
    \begin{tabular}{ccccccc}
        \hline
        & \multicolumn{2}{c}{Siemens Test Dataset} & \multicolumn{2}{c}{GE Test Dataset} & \multicolumn{2}{c}{ACR CT Phantom}\\
        & \multicolumn{2}{c}{$\mathrm{CTDI_{vol}}$: 1.9-7.2 mGy} & \multicolumn{2}{c}{$\mathrm{CTDI_{vol}}$: 2.3-5.4 mGy} & \multicolumn{2}{c}{$\mathrm{CTDI_{vol}}$: 3.4 mGy}\\
        & \multicolumn{2}{c}{WL: 40, WW: 300} & \multicolumn{2}{c}{WL: 40, WW: 400} & \multicolumn{2}{c}{Full-range}\\
        \cmidrule(lr){2-3}\cmidrule(lr){4-5}\cmidrule(lr){6-7}
        & MSE$\downarrow$ & SSIM$\uparrow$ & MSE$\downarrow$ & SSIM$\uparrow$ & MSE$\downarrow$ & SSIM$\uparrow$\\
        \hline
        Baseline & $774.46\pm811.18$ & $0.9577\pm0.0323$ & $1768.27\pm1040.13$ & $0.9152\pm0.0389$ & $187.16\pm26.89$ & $0.9842\pm0.0016$\\
        BM3D & $670.30\pm755.00$ & $0.9631\pm0.0296$ & $1176.68\pm902.82$ & $0.9424\pm0.0339$ & $184.61\pm26.52$ & $0.9844\pm0.0016$\\
        NLM & $724.50\pm780.64$ & $0.9605\pm0.0304$ & $1268.01\pm944.06$ & $0.9391\pm0.0347$ & $187.09\pm26.86$ & $0.9842\pm0.0016$\\
        RED-CNN & $210.50\pm208.29$ & $0.9863\pm0.0122$ & $477.36\pm216.44$ & $0.9713\pm0.0126$ & $77.40\pm20.93$ & $0.9925\pm0.0018$\\
        WGAN & $453.66\pm313.47$ & $0.9691\pm0.0168$ & $1004.27\pm473.89$ & $0.9449\pm0.0236$ & $179.32\pm122.46$ & $0.9874\pm0.0025$\\
        CPCE-3D & $324.67\pm284.80$ & $0.9797\pm0.0156$ & $761.76\pm368.73$ & $0.9589\pm0.0184$ & $130.91\pm96.21$ & $0.9906\pm0.0025$\\
        QAE & $241.36\pm274.95$ & $0.9853\pm0.0134$ &  $526.67\pm236.40$ & $0.9699\pm0.0132$ & $87.41\pm23.91$ & $0.9913\pm0.0024$\\
        DP-ResNet & $270.77\pm122.73$ & $0.9795\pm0.0080$ & $451.32\pm202.43$ & $0.9734\pm0.0117$ & $153.08\pm193.50$ & $0.9927\pm0.0033$\\
        EDCNN & $240.39\pm275.13$ & $0.9848\pm0.0144$ &  $523.12\pm253.46$ & $0.9694\pm0.0139$ & $84.48\pm18.74$ & $0.9919\pm0.0019$\\
        TransCT & $250.96\pm226.68$ & $0.9836\pm0.0141$ & $512.06\pm221.15$ & $0.9700\pm0.0126$ & $98.43\pm72.26$ & $0.9915\pm0.0026$\\
        DU-GAN & $286.49\pm309.19$ & $0.9820\pm0.0162$ & $584.49\pm275.07$ & $0.9667\pm0.0150$ & $108.70\pm23.07$ & $0.9898\pm0.0018$\\
        CTformer & $242.06\pm282.92$ & $0.9847\pm0.0147$ & $785.66\pm501.63$ & $0.9550\pm0.0221$ & $80.16\pm19.46$ & $0.9925\pm0.0015$\\
        Ours & $\textbf{184.98}\pm\textbf{106.53}$ & $\textbf{0.9881}\pm\textbf{0.0067}$ & $\textbf{422.39}\pm\textbf{201.62}$ & $\textbf{0.9751}\pm\textbf{0.0115}$ & $\textbf{51.62}\pm\textbf{14.67}$ & $\textbf{0.9951}\pm\textbf{0.0014}$\\
        \hline
    \end{tabular}
\end{table*}

\begin{figure*}[htb]
    \centering
    \includegraphics[scale=1]{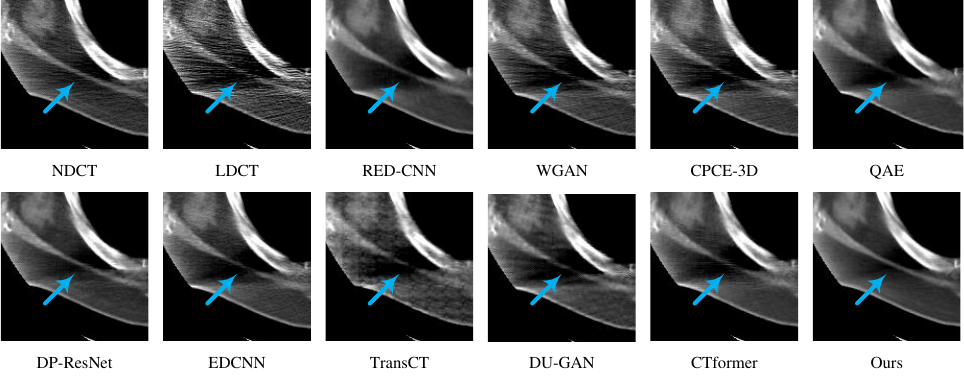}
    \caption{Visualization of a selected region with high contrast. The blue arrow indicates a continuous high-density structure (Slice ID: Siemens L266-002, WL: 40, WW: 300).}
    \label{fig-visobj1}
\end{figure*}

\subsection{Objective Evaluation}
We conducted an objective evaluation to compare the proposed framework with nine state-of-the-art learning-based methods, namely RED-CNN \cite{chen2017low}, WGAN \cite{yang2018low}, CPCE-3D \cite{shan20183}, QAE \cite{fan2019quadratic}, DP-ResNet \cite{yin2019domain}, EDCNN \cite{liang2020edcnn}, TransCT \cite{zhang2021transct}, DU-GAN \cite{huang2021gan}, and CTformer \cite{wang2022ctformer}. We retrained all these methods using our training dataset for a fair comparison. Specifically, as the overall volume of the training dataset differs from their original setups, we retrained each model with more iterations. The retraining was terminated when the validation performance saturated, which also applied to GAN-based methods as they still employed pixel-wise paired supervision such as L1 or L2. The performance of two traditional denoising methods, i.e., BM3D \cite{dabov2007image} and non-local means (NLM) \cite{buades2005non}, was also tested as references. We employed MSE and SSIM as quantitative evaluation metrics. The results are reported in Table \ref{table-objeval1}.

\begin{figure*}[htb]
    \centering
    \includegraphics[scale=1]{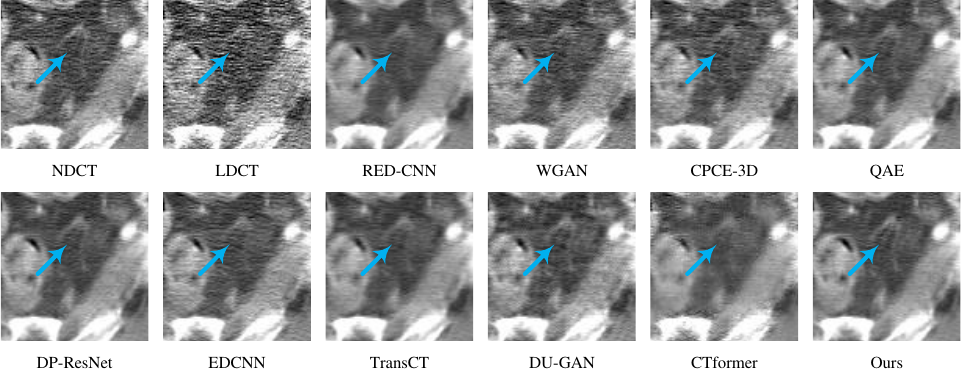}
    \caption{Visualization of a selected region with severe streak artifacts. The blue arrow indicates a low-contrast fine structure (Slice ID: GE L235-097, WL: 40, WW: 300).}
    \label{fig-visobj2}
\end{figure*}

Our method reduces the MSE score by up to 70\%, outperforming the others by a significant margin in all studies, indicating its superiority. Furthermore, our results on two scanners show high consistency, whereas some methods, such as DP-ResNet, EDCNN, TransCT, and CTformer, witness lower robustness. To better visualize the image quality, we present two sample slices in their corresponding zoom-in patches: In Fig. \ref{fig-visobj1}, the proposed method successfully recovers the continuous structure; in comparison, most methods failed to remove the heavy streak artifacts. In Fig. \ref{fig-visobj2}, the branches of the low-contrast structure have better visibility and sharpness in our results. Although the results from DP-ResNet are also promising, a certain degree of blurriness is observed.

\subsection{Subjective Test}
Although the quantitative evaluation shows significant improvement of the proposed framework against other methods, the evaluation metrics (i.e., MSE, and SSIM) might not reflect the real-world application scenario. In other words, a professional radiologist may pay special attention to certain aspects rather than general image quality metrics when reviewing CT images. As the main application is to assist clinical diagnosis, two radiologists with clinical experience of 12 years and 9 years were invited to perform a series of subjective evaluations.

\begin{table}[htb]
    \renewcommand{\arraystretch}{1.3}
    \caption{Results of Lesion Detection}
    \label{table-subeval1}
    \centering
    \begin{tabular}{ccccc}
        \hline
        Method & Recall & Precision & Accuracy & F1\\
        \hline
        RED-CNN & 0.5128 & 0.5714 & 0.3929 & 0.5405\\
        WGAN & 0.4250 & 0.8947 & 0.4444 & 0.5763\\
        CPCE-3D & 0.4390 & 0.8182 & 0.4375 & 0.5714\\
        QAE & 0.4595 & 0.6296 & 0.4000 & 0.5312\\
        DP-ResNet & 0.5000 & 0.7407 & 0.4375 & 0.5970\\
        EDCNN & 0.5122 & 0.7500 & 0.4375 & 0.6087\\
        TransCT & 0.4872 & 0.7037 & 0.4510 & 0.5758\\
        DU-GAN & 0.3500 & 0.6667 & 0.3400 & 0.4590\\
        Ctformer & 0.5000 & 0.8400 & 0.5000 & 0.6269\\
        Ours & \textbf{0.6098} & \textbf{0.9615} & \textbf{0.6304} & \textbf{0.7463}\\
        \hline
    \end{tabular}
\end{table}

The first subjective evaluation was performed using the 2016 Low-dose CT AAPM Grand Challenge test dataset, which consists of 16 patient scans with lesions and 4 healthy references. The radiologists were asked to mark all the lesions without prior knowledge of the patient's diagnosis. By comparing the radiologists' annotations with the ground truth provided by the challenge committee, we calculated precision, recall, accuracy, and F1 score for each method. The results are listed in Table \ref{table-subeval1}. All the metrics of the proposed framework show significant improvements compared to the others. On the one hand, our result receives the highest precision over 0.9, meaning that it does not produce artifacts that could affect the diagnostic acceptability. On the other hand, the higher recall rate indicates better diagnostic sensitivity than the others. Overall measurements in precision and the F1 score further confirm the true value of the proposed framework in clinical exams.

\begin{table*}[htb]
    \renewcommand{\arraystretch}{1.3}
    \caption{Results of Subjective Quality Evaluation}
    \label{table-sujeval2}
    \centering
    \begin{tabular}{ccccccccccc}
        \hline
        & RED-CNN & WGAN & CPCE-3D & QAE & DP-ResNet & EDCNN & TransCT & DU-GAN & CTformer & Ours\\
        \hline
        Noise suppression & 4.38 & 3.08 & 3.19 & 3.65 & 4.38 & 4.15 & 4.35 & 3.31 & 3.69 & \textbf{4.62}\\
        Contrast retention & 3.88 & 3.77 & 3.96 & 3.73 & 3.42 & 4.15 & 3.62 & 4.12 & 3.85 & \textbf{4.46}\\
        Margin sharpness & 2.65 & 2.85 & 2.81 & 2.58 & 2.46 & 2.73 & 2.46 & 2.69 & 2.54 & \textbf{3.38}\\
        Diagnostic acceptability & 3.58 & 3.31 & 3.58 & 3.31 & 3.73 & 3.81 & 3.50 & 3.46 & 3.35 & \textbf{4.19}\\
        \hline
    \end{tabular}
\end{table*}

The second subjective evaluation was to evaluate the perceptual quality of each method using the Siemens test dataset. During the test, three images, including a normal-dose CT, a low-dose CT, and a refined low-dose CT, were presented to the reviewer. For each study, the reviewers were requested to assign scores concerning noise suppression, contrast retention, margin sharpness, and diagnostic acceptability\footnote{All studies in the Siemens dataset contain at least one lesion.}, respectively. A five-point scale table was employed, where the lowest score (1) was assigned to low-dose CT whereas the highest score (5) was assigned to full-dose CT. All processed studies from each method were de-identified and shuffled to avoid biased judgments. Since there were no absolute clean references and the full-dose counterparts still contained a certain amount of noise, the radiologists were asked to assign scores no higher than 5, no matter whether the sample was of higher quality than the full-dose reference. The results are summarized in Table \ref{table-sujeval2}. Again, the proposed method receives the highest scores by a significant margin in all aspects, which is well-aligned with the results of the previous evaluations.

\begin{figure}[htb]
    \centering
    \includegraphics[scale=1]{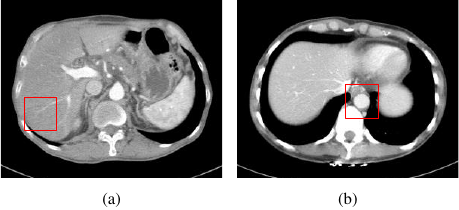}
    \caption{Illustration of the two samples, where the box regions in red were identified as (a) metastasis pancreas and (b) aorta. Their zoom-in comparisons are presented in Figs. \ref{fig-visex1}-\ref{fig-visex2}. (WL: 10, WW: 400, source: the AAPM Grand Challenge committee)}
    \label{fig-visall}
\end{figure}

\begin{figure*}[b]
    \centering
    \includegraphics[scale=1]{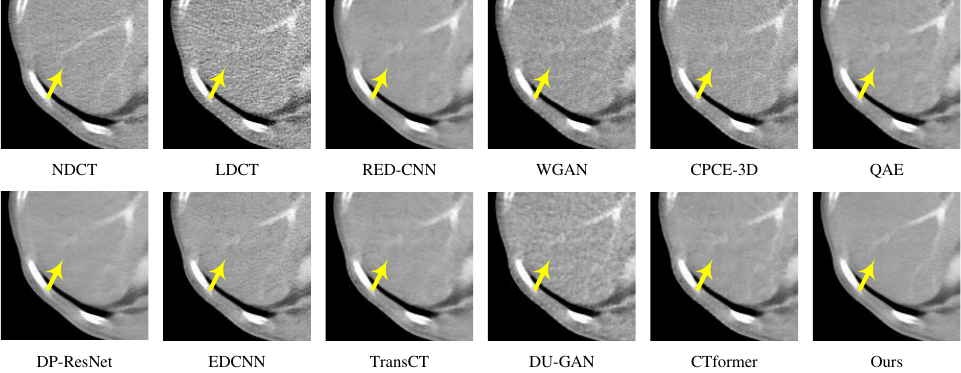}
    \caption{Zoom-in visualization of the red box region in Fig. \ref{fig-visall}(a). The low-contrast lesion is marked by the yellow arrow (Slice ID: L593-050, WL: 10, WW: 400, source of diagnosis: the AAPM Grand Challenge committee).}
    \label{fig-visex1}
\end{figure*}

\begin{figure*}[t]
    \centering
    \includegraphics[scale=1]{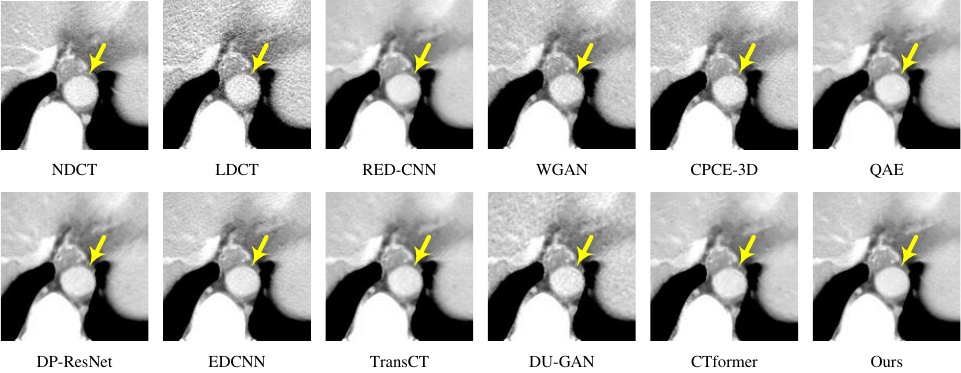}
    \caption{Zoom-in visualization of the red box region in Fig. \ref{fig-visall}(b). The yellow arrow marks the uniformly distributed aorta (Slice ID: L548-050, WL: 10, WW: 300).}
    \label{fig-visex2}
\end{figure*}

Interestingly, RED-CNN, DP-ResNet, and TransCT get high noise suppression scores but low margin sharpness scores due to aggressive denoising (i.e., over-smoothing). On the contrary, introducing detail preservation constraints (e.g., perceptual loss in CPCE-3D and adversarial loss in WGAN and DU-GAN) leads to compromised denoising performance and reduced diagnostic sensitivity. In comparison, the proposed method can reach a pleasing balance between these aspects, yielding satisfactory results that radiologists prefer.

Last, we show two representative examples from these evaluations: Fig. \ref{fig-visall} presents an overview of the two samples in full-dose, in which their zoom-in crops are compared in Figs. \ref{fig-visex1}-\ref{fig-visex2}: Fig. \ref{fig-visex1} depicts a lesion that is barely noticeable by experts due to its low contrast. Similar to the full-dose references, our method could transfer details more faithfully and maintain the contrast but with an even lower noise level. In comparison, some methods either fail to suppress noise (e.g., WGAN, DU-GAN) or introduce artifacts (e.g., CTformer), making them less reliable in clinical exams. Fig. \ref{fig-visex2} visualizes the cross-section of the aorta, where the higher uniformity in our result indicates better denoising quality.

\begin{figure*}[htb]
    \centering
    \includegraphics[scale=1]{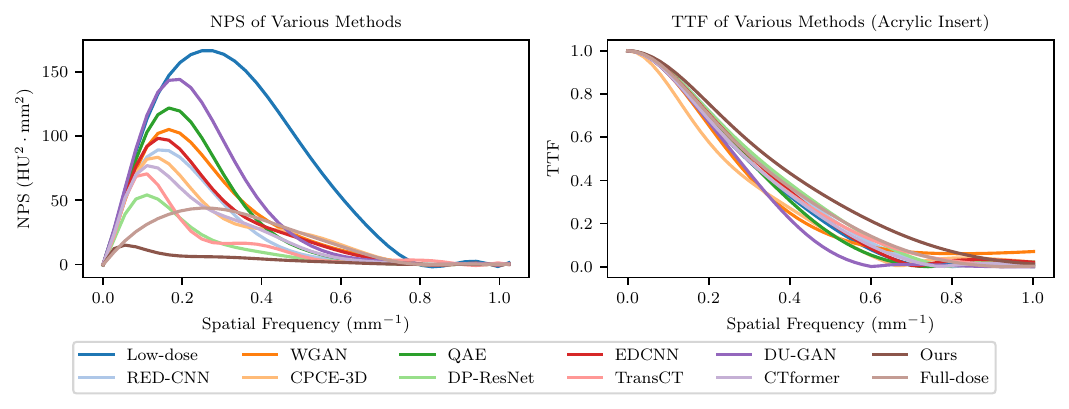}
    \caption{NPS and TTF results of learning-based methods on ACR CT phantom.}
    \label{fig-nps-ttf-all}
\end{figure*}

\begin{figure*}[htb]
    \centering
    \includegraphics[scale=1]{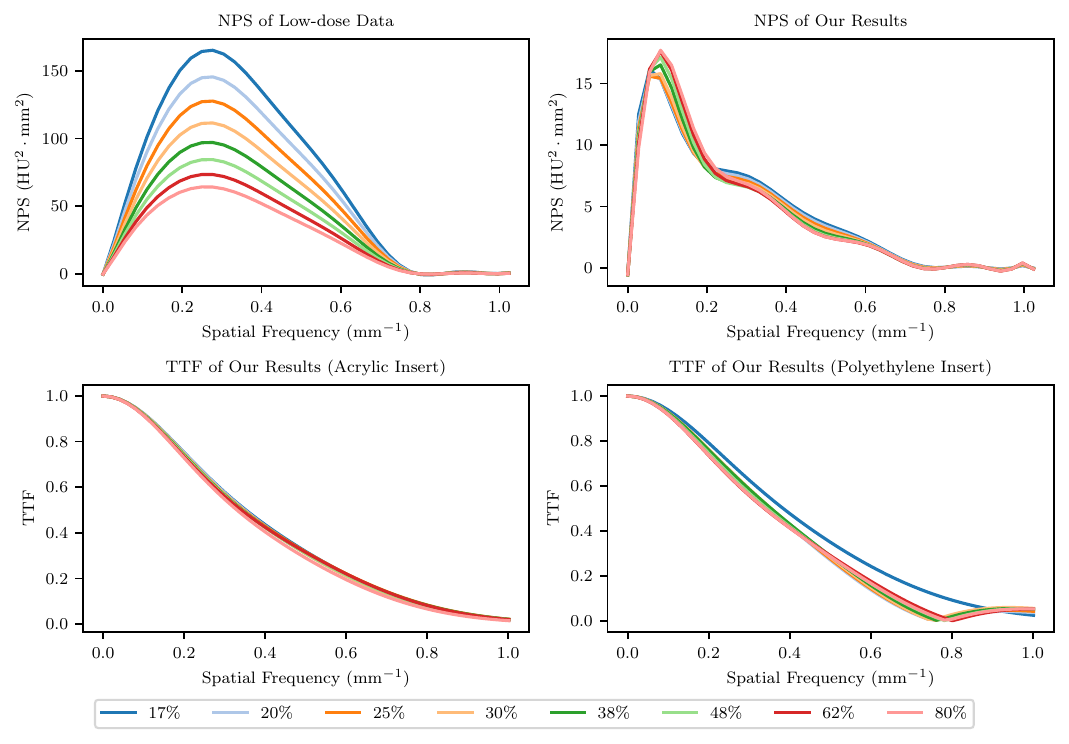}
    \caption{NPS and TTF results of our method on various simulated dose levels on ACR CT phantom.}
    \label{fig-nps-ttf-ours}
\end{figure*}

\subsection{Phantom Examination}
\label{sec:phantom}
To further analyze the standardized CT properties, we conducted an examination using ACR 464 CT Phantom. In this experiment, two real scans representing the full dose (13.5 mGy) and the quarter dose (3.4 mGy) of the same phantom were obtained. Because the same scanner model was also used previously (i.e., Siemens SOMATOM Definition Flash), we directly applied all the learning-based models to these scans without retraining. The results in terms of MSE and SSIM are reported in Table \ref{table-objeval1}, where the proposed method trained using synthetic data still works well in real clinic exams. After that, the slices of the phantom's first and third layers were used to analyze the task-based transfer function (TTF) and NPS. The methods and default settings from \cite{greffier2022iqmetrix} were employed to obtain the results.

First, we show the NPS plot and TTF plot of each method in Fig. \ref{fig-nps-ttf-all}. It can be observed that every candidate achieves a certain degree of noise reduction on the low-dose phantom scan. However, many do not show satisfactory noise suppression on low-frequencies. The facts of lower peak frequency, higher noise magnitude, and degraded transfer function indicate increased blurriness and decreased spatial resolution in the resulting images. In contrast, the proposed method produces a much smoother NPS with a lower degree of noise than the full-dose version. Although the noise power at the low-frequency band is slightly higher than the full-dose reference, its improved TTF implies that the details are reproduced more faithfully, whereas sharpness degradation was marginal.

Second, we analyzed the robustness under various dose levels. Following the noise insertion method described in \cite{zeng2022performance}, we generated a bunch of simulated low-dose projections ranging from 17\% to 80\% of the full dose level. Fig. \ref{fig-nps-ttf-ours} presents the dose-dependent NPS and TTF plots. The resultant NPS plots of the proposed method show consistent denoising performance across a wide range of dose levels, and the TTF plots of two inserts further support its high robustness.

\begin{table}[htb]
    \renewcommand{\arraystretch}{1.3}
    \centering
    \begin{threeparttable}
        \caption{CT Number Accuracy under Various Dose Levels}
        \label{table-ctnumber}
        \begin{tabular}{ccccc}
        \hline
        \multirow{2}{*}{Dose Level} & Air & Bone & Acrylic & Polyethylene\\
        & (-1000 HU) & (955 HU) & (120 HU) & (-95 HU)\\
        \hline
        17\% & -968.31 & 847.61 & 116.65 & -84.84\\
        20\% & -968.28 & 847.80 & 116.67 & -84.85\\
        25\% & -968.26 & 847.96 & 116.68 & -84.85\\
        30\% & -968.24 & 848.01 & 116.70 & -84.84\\
        38\% & -968.25 & 847.97 & 116.71 & -84.83\\
        48\% & -968.25 & 847.90 & 116.69 & -84.80\\
        62\% & -968.25 & 847.88 & 116.67 & -84.77\\
        80\% & -968.24 & 847.91 & 116.65 & -84.73\\
        100\%\tnote{a} & -968.38 & 847.34 & 116.60 & -84.69\\
        \hline
        \end{tabular}
        \begin{tablenotes}
            \item[a] Reference full-dose images without denoising.
        \end{tablenotes}
    \end{threeparttable}
\end{table}

Finally, we measured the CT number accuracy of the processed images against the full-dose references. The results are listed in Table \ref{table-ctnumber}. These reports confirm that no particular bias is presented in the reconstructed images, and dose levels do not affect this observation.

\section{Conclusions}
\label{sec:conclusions}
The main goal of this paper is to improve the quality of LDCT images for multi-slice spiral CT. We comprehensively discussed the proposed two-stage processing pipeline across both projection and image domains. We analyzed the impacts of rebinning and filtered back-projection to the final reconstruction. To fully utilize the intra-slice and inter-slice similarity inherent in the acquired projection volume, we transformed the task into a multi-frame-based denoising and refinement problem. Although conventional rebinning and reconstruction methods link the two networks, we performed several studies to verify the effectiveness of our method and show its superiority over state-of-the-art methods through qualitative and quantitative performance evaluations. Future work might include better detail preservation via higher-level loss functions other than a simple L1 penalty.

\section*{Acknowledgment}
The authors would like to thank the anonymous reviewers for their valuable feedback and constructive suggestions, which have greatly improved the quality and clarity of this manuscript.

\bibliographystyle{IEEEtran}
\bibliography{IEEEabrv,IEEEexample}

\end{document}